\documentclass[aps,pra,showpacs,twocolumn]{revtex4}

\usepackage{graphicx}
\usepackage{amsmath}
\begin{document}
\title{Towards a theory of attosecond transient recorder}  
\author{E.E. Krasovskii$^{1,2}$ and M.~Bonitz$^{3}$\medskip}

\affiliation{
$^1$Departamento de F\'isica de Materiales, Facultad de Qu\'imicas,
    Universidad del Pa\'is Vasco, San Sebasti\'an/Donostia, Basque Country, Spain\\
$^2$Donostia International Physics Center (DIPC), San Sebasti\'an/Donostia, 
    Basque Country, Spain\\
$^3$Institut f\"ur Theoretische Physik und Astrophysik,
Universit\"at Kiel, D-24098 Kiel, Germany
}

\begin{abstract}
Laser assisted photoemission by a chirped subfemtosecond extreme ultraviolet (XUV) 
pulse is considered within an exactly solvable quantum-mechanical model. Special 
emphasis is given to the energy dependence of photoexcitation cross-section. The 
streaked spectra are analyzed within the classical picture of initial time-momentum 
distribution $r_{\rm ini}(p,t)$ of photoelectrons mapped to the final energy scale. 
The actual time-momentum  distribution in the absence of the probe laser field is 
shown to be a poor choice for $r_{\rm ini}$, and a more adequate ansatz is suggested. 
The semiclassical theory offers a simple practically useful approximation for streaked 
spectra. Its limitations for sufficiently long chirped XUV pulses are established. 
\end{abstract} 
\maketitle 

\section{introduction}
Subfemtosecond x-ray pulses are becoming a powerful tool for studying ultrafast 
processes in atoms~\cite{Hen2001N,Dre2001S,Kie2004N} and solids~\cite{Cav2007N}, 
and their range of applications is growing rapidly~\cite{KI09}. The characterization 
of ultrashort pulses is an important problem in this field. The attosecond metrology 
is based on the laser assisted XUV (extreme ultraviolet) photoemission, whereby 
the temporal structure of the XUV pulse is reflected in the resulting streaked 
photoelectron spectrum~\cite{Hen2001N,Dre2001S}. A subfemtosecond XUV pulse creates 
a photoelectron wave packet temporally confined within a fraction of the oscillation 
period $T_{\rm L}$ of the laser light. The wave packet is accelerated or decelerated
by the superimposed laser field, and the dependence of the kinetic energy of the 
photoelectrons on the release time allows to record the time dependent spectra with 
a resolution around 100~as~\cite{Kie2004N}. 

The gross features of the process can be understood within a quasi-classical model, 
in which the evolution of the photoelectron wave packet is split into two steps: 
first, a short XUV pulse creates an initial momentum distribution, which is then 
treated as a distribution of classical particles accelerated by the laser electric 
field. The model explains modulations of the width and center of gravity of the 
photoelectron spectra depending on the time delay of the laser pulse relative to 
the XUV pulse~\cite{Hen2001N,Dre2001S}. Quantum-mechanical calculations of laser 
assisted atomic photoionization performed in Refs.~\cite{itatani02,kitzler02} have 
confirmed that the XUV pulse duration can be measured by the width modulation. 
Recently, various quantum-mechanical formalisms have been proposed that reveal the 
connection between temporal characteristics of the pump pulse and the streaked 
spectrum and form a basis for characterizing both the pulse and the electronic 
system being excited~\cite{KK06,YPCB07}. Because the pump perturbation cannot be 
completely disentangled from the probe laser field the inverse problem -- the 
reconstruction of the input pulse from the output spectra -- is a difficult one. 

In Ref.~\cite{Kie2004N} an XUV pulse of duration 250~as was first measured 
by the streaked photoelectron spectra from neon atoms, and it has been 
shown that the streaked spectrum can be treated as a tomographic image of the 
{\em initial time-momentum distribution of photoelectrons}. The two-step model has 
become a common paradigm in laser assisted photoemission~\cite{Gou2007S,KV08ARPC,QMI05}. 
Although the initial time-momentum distribution $r_{\rm ini}(p,t)$ appears to be an 
important concept in the theory of photoelectron streaking it is not straightforward 
to rigorously define this notion. A quantum-mechanical derivation of $r_{\rm ini}$ 
performed in Ref.~\cite{YBS05} showed that under certain assumptions a function 
independent of the streaking field can indeed be constructed, and it reduces to 
the time-momentum distribution $r_{\rm 0}(p,t)$ in the absence of the laser field. 
In Ref.~\cite{KB07} an alternative prescription for $r_{\rm ini}$ was introduced, 
which very satisfactorily described the streaked photoemission lineshape for a wide 
range of the XUV pulse durations and delay times. Understanding of the physical 
meaning and usage of the function $r_{\rm ini}(p,t)$ is important in order to 
correctly interpret the streaked spectra and obtain information about ultrafast 
electronic processes.

In the present paper the classical approach based on $r_{\rm ini}(p,t)$, 
which hitherto has been used mainly for illustrative purposes, will be 
shown to have predictive power. We shall develop the semiclassical 
superposition approximation (SSA), in which the pump and the probe 
pulses can be disentangled. For a bandwidth-limited pump pulse it 
reduces the calculation of the final momentum intensity distribution 
of photoelectrons $J(p)$ to the integral~\cite{KB07}
\[
J(p)=\int_{-\infty}
         ^{+\infty}U(t) I[p-\Delta p(t)] S[p-\Delta p(t)]\;dt,
\]
where the shape of the pump pulse enters through its time intensity
envelope $U(t)$ and spectral intensity envelope $I(p)$, and the properties 
of the system enter through the excitation cross-section $S(p)$. The 
effect of the laser field is described by the momentum $\Delta p(t)$ 
transferred to the electron created at the time moment $t$. The choice 
of the integrand will be explained in Sec.~\ref{LaserDressedNoChirp_A}, 
and in Sec.~\ref{LaserDressedChirpedPulse} it will be extended to chirped
pulses by replacing the function $I(p)$ in the integrand by a function
$g(p,t)$ with an explicit time dependence:
\[
J(p)=\int_{-\infty}
         ^{+\infty}U(t) g[p-\Delta p(t),t] S[p-\Delta p(t)]\;dt.
\] 
We shall consider a one-dimensional prototype of the photoexcitation of an
atom or a localized state at a solid state surface and apply the SSA to link 
the input to the output using the initial time-momentum distribution as an 
intermediary. The paper is organized as follows. The computational method is 
described in Sec.~\ref{method}. In Sec.~\ref{LaserFreeCase} we analyze the 
time-momentum distribution of photoelectrons from a chirped pulse in the 
absence of the laser field. Section~\ref{LaserDressedNoChirp} discusses laser 
dressed photoemission by non-chirped pulses: it reveals problems in defining 
the initial time-momentum distribution function $r_{\rm ini}$ and introduces a 
practical ansatz for $r_{\rm ini}$. Section~\ref{LaserDressedChirpedPulse}  
generalizes SSA to chirped pulses and discusses the inverse problem. 

\section{model parameters and method}\label{method}
\begin{figure}[t]
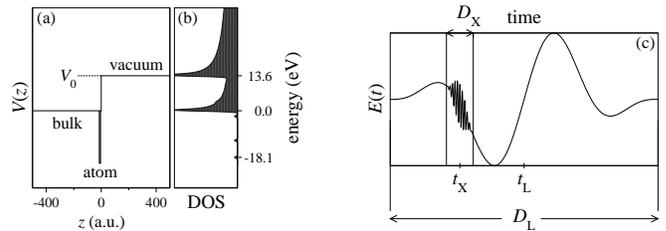
   
\begin{minipage}{0.48\textwidth}
\includegraphics[width=0.45\textwidth]{fig1a.eps}
\hfill 
\includegraphics[width=0.45\textwidth]{fig1c.eps}
\end{minipage}
\caption{\label{box} 
(a) Potential $V(z)$ of the model 1D system. The atom is represented by a 
6~a.u. wide potential well centered at $z=-8$~a.u. relative to the surface 
barrier. (b) Density of states. (c) Superposition of the electric field of 
the XUV pulse of duration $D_{\textsc x}$ and the laser pulse of duration 
$D_{\textsc l}$.
}
\end{figure}

In order to avoid any numerical approximations in solving the time-dependent 
Schr\"odinger equation we employ the one-dimensional model adopted in 
Ref.~\cite{KB07}. The unperturbed Hamiltonian (in the units $\hbar=1$, $m=1/2$) 
is $\hat H=-\Delta+V(z)$. The model potential $V(z)$ is a piecewise constant 
function shown in Fig.~\ref{box}(a). The system is enclosed in a box, so that 
the spectrum of $\hat H$ is discrete. This setup describes the photoemission 
from a localized atomic state in a crystal close to the surface. We shall 
consider the lowest state of the model atom [$\epsilon_{\rm i}=-18.1$~eV, 
see Fig.~\ref{box}(b) and the figure caption]. The XUV pulse creates two 
wave packets traveling in opposite directions. We shall consider the spectrum
of the one traveling to the right. Although the presence of the surface barrier 
between bulk and vacuum, see Fig.~\ref{box}(a), is not essential for the present 
application, the barrier is retained for technical convenience: it reduces the 
total number of energy levels and slows down the right traveling packet. 

The electric field $E(t)$ applied to the system is a superposition of
the XUV (subscript X) and laser (L) pulses:
\begin{equation}\nonumber
E(t)=E_{\textsc x}(t-t_{\textsc x})\cos[\omega_{\textsc x}(t-t_{\textsc x})] 
    +E_{\textsc l}(t-t_{\textsc l})\sin[\omega_{\textsc l}(t-t_{\textsc l})].
\end{equation}
The XUV and the laser pulse are confined to the intervals of length $D_{\textsc{x}}$ 
and $D_{\textsc{l}}$ centered at $t_{\textsc x}$ and $t_{\textsc l}$, respectively, and 
their envelopes $E_{\textsc x}$ and $E_{\textsc l}$ have the same shape
\begin{equation}\nonumber
E_{\textsc{x,l}}(t)=E_{\textsc{x,l}}^0\cos^2\frac{\pi t}{D_{\textsc{x,l}}}
\;\;
{\rm for}
\;\; t\in\left[-\frac{D_{\textsc{x,l}}}{2},
                               \frac{D_{\textsc{x,l}}}{2}\right], 
\end{equation}
see Fig.~\ref{box}(c). Note that $D_{\textsc{x,l}}$ are twice the FWHM duration 
often used in the literature. The intensity spectrum $I(\omega)$ of such XUV 
pulse is close to Gaussian. To construct the classical model we shall need 
the normalized temporal intensity envelope of the XUV pulse $U(t)$ defined by
the formula
\begin{equation}\label{u_envelope}
U(t)=\left[E_{\textsc x}(t)/E_{\textsc x}^0\right]^2. 
\end{equation}

All calculations are performed for $\omega_{\textsc x}=91$~eV and 
$E_{\textsc x}^0=10^6$~V/cm, which is within the perturbational limit. 
We shall also consider linearly chirped pulses, with $\omega_{\textsc x}$ 
changing in time: 
\begin{equation}\label{chirp_formula}
\omega_{\textsc x}(t) = \omega_{\textsc x}^0 + b(t-t_{\textsc x}).
\end{equation}
The parameters of the laser pulse applied to the system are typical of 
attosecond spectroscopy~\cite{Gou2004S}: all calculations are performed 
for frequency $\omega_{\textsc l}=1.65$~eV ($T_{\textsc l}=2.5$~fs) and 
duration $D_{\textsc l}=5$~fs. In the following we set the center of the
laser pulse to zero, $t_{\textsc l}=0$. If not otherwise stated the laser 
field  amplitude is $E_{\textsc l}^0=6\times 10^7$~V/cm.  

The perturbation in dipole approximation is $zeE(t)$, and the time-dependent 
Schr\"odinger equation (TDSE) is solved in matrix form in terms of exact 
eigenfunctions of $\hat H$. In spite of its simplicity the model is quite 
realistic. In particular, the energy variations of the excitation cross-section 
are in accord with those experimentally observed, see Ref.~\cite{KB07}. A more 
sophisticated development of the model has been recently applied to laser 
assisted photoemission from solids~\cite{KazE09}. The wave packets created
by the XUV pulse can be driven back to the ion and rescatter, thereby absorbing 
additional UV photons. This was studied for ionization in Ref.~\cite{BB08} 
and for electron scattering on an ion in Ref.~\cite{BB09} by solving the TDSE 
for a 1D and 2D model system. In the present work we are interested in the 
energy distribution $J(\epsilon)$ of photoelectrons recorded by a detector 
far away from the excitation region. The spectrum $J(\epsilon)$ is calculated 
by reexpanding the packet moving to the right in terms of the eigenfunctions 
of the system.

We shall adopt the following convention about notation. The three
variables -- energy $\epsilon$, momentum $p$, and frequency (XUV
photon energy) $\omega$ -- can be used one instead of another as
argument of frequently used functions. They are simply connected:
$\omega=\epsilon-\epsilon_{\rm i}$, with $\epsilon_{\rm i}$ being the
energy of the photoemission initial state; $\epsilon=p^2+V_0$. The
kinetic energy of the photoelectron in vacuum is, thus, 
$\epsilon_{\rm i}+\omega-V_0$.  For the sake of brevity, we shall use 
$\epsilon$, $p$ or $\omega$ depending on the context, e.g., $J(\omega)$ 
is the short for $J(\omega+\epsilon_{\rm i})$, and $J(p)$ is the short for
$J(p^2+V_0)$.

\section{XUV emission without laser field}\label{LaserFreeCase}
\begin{figure}[t] 
\includegraphics[width=0.48\textwidth]{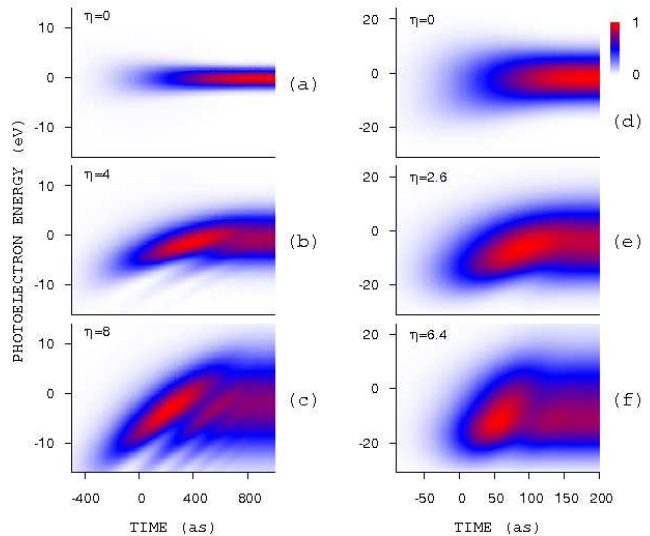}
\caption{\label{MAP1} (color online)
Time dependent occupation of final states $J(\epsilon,t)$ for the duration 
$D_{\textsc x}=2000$~as, $\Gamma_{\textsc x}^0=3$~eV (left column) and 400~as, 
$\Gamma_{\textsc x}^0=15$~eV (right column). The chirp rate $b$ is zero in 
graphs (a) and (d), and $b=6$~(b), 12~(c), 96~(e), and 240~eV/fs~(f). The 
photoelectron energy is given relative to $\epsilon_{\rm i}+\omega_{\textsc x}^0$ 
[the frequency at $t=0$ is $\omega_{\textsc x}^0=91$~eV, see 
Eq.~(\ref{chirp_formula})] and time relative to $t_{\textsc x}$. The maximum 
intensity is set to 1. The vertical cross section at the right border of the 
map is the final intensity distribution $J(\epsilon)$ recorded by the detector.
}
\end{figure}

The temporal evolution of the photoelectron spectrum during the XUV pulse is 
shown in Fig.~\ref{MAP1} for a bandwidth limited and for linearly chirped 
pulses. The color map shows the time and energy dependent occupation number 
$J(\epsilon,t)$ of eigenstates $|\,\epsilon\,\rangle$. The vertical cross 
section of the map at the end of the pulse 
$J(\epsilon,t_{\textsc x}+\frac{1}{2}D_{\textsc x})\equiv J(\epsilon)$ is the 
energy distribution curve (EDC) measured in the experiment. The time 
derivative of the function $J(\epsilon,t)$ gives the energy resolved 
transition rate 
\begin{equation}\label{r0}
r_0(\epsilon,t) = dJ(\epsilon,t)/dt.
\end{equation}
The function $r_0(p,t)$ is, thus, the time-momentum distribution of 
photoelectrons in the absence of the laser field.

In the absence of the laser field the observed final EDC obeys the
simple formula~\cite{KB07}:
\begin{equation}\label{simple}
J(\epsilon)\sim I(\epsilon)S(\epsilon),
\end{equation}
where $I(\omega)$ is the spectral intensity distribution of the XUV pulse, 
and $S(\epsilon)$ is the photoemission cross-section,
\begin{equation}\label{cross-section}
S(\epsilon)=
\left |\,\langle\,\epsilon\,|\,z\,|\,\epsilon_{\rm i}\,\rangle\,\right |^2.
\end{equation}
Equation~(\ref{simple}) predicts the photoelectron spectrum without solving the 
TDSE. It is a consequence of the weak XUV perturbation and of the negligible
probability of transitions within the (quasi-) continuum, $\epsilon>0$, see 
Appendix. (This relation is also obtained in the perturbational limit of the 
approximation to the TDSE introduced by Lewenstein {\it et al.}~\cite{LBI94}, 
i.e., in the limit when the vector potential of the electric field can be neglected 
compared to the electron momentum.) For the present system it is found to hold 
for XUV pulses over a wide range of pulse parameters. Also the entire map 
$J(\epsilon,t)$ calculated from Eq.~(\ref{simple}) practically coincides with 
the exact result.

The spectrum evolution is seen to strongly depend on the chirp rate. The 
effect of a linear chirp on the intensity spectrum can be characterized 
by the dimensionless index $\eta=bD_{\textsc x}/\Gamma_{\textsc x}^0$, where 
$\Gamma_{\textsc x}^0$ is the full spectral width at half maximum (FWHM) of 
the non-chirped pulse with the same envelope and the carrier frequency 
$\omega_{\textsc x}^0$. When $\eta$ approaches 4 the $J(\epsilon,t)$ map 
starts acquiring a characteristic stripe structure, see Fig.~\ref{MAP1}(b,c,f). 
However, it is not straightforward to determine the value of the chirp rate 
from a known $J(\epsilon,t)$ distribution. 

\begin{figure}[b] 
\includegraphics[width=0.45\textwidth]{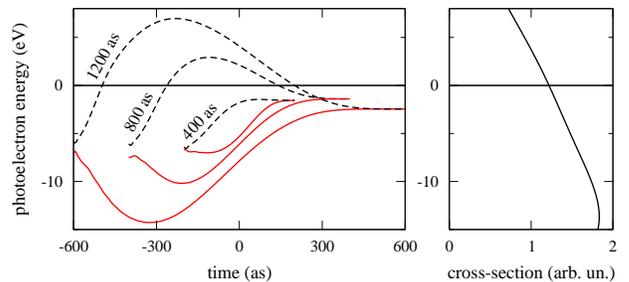}
\caption{\label{f3} (color online)
Time dependence of the photoelectron spectrum center of gravity 
$\epsilon_g(t)$ for the chirp rate $b=\pm 24$~eV/fs for three pulse 
durations: $D_{\textsc x}=400$, 800, and 1200~as. Positive (negative) 
chirp is shown by full (dashed) lines. Right panel shows the function 
$S(\epsilon)$, see Eq.~(\ref{cross-section}). Photoelectron energy 
$\epsilon$ is given relative to $\epsilon_{\rm i}+\omega_{\textsc x}^0$.
}
\end{figure} 
To illustrate the point, we show in Fig.~\ref{f3} the time dependence of the 
center of gravity of the EDC $\epsilon_g(t)$ obtained by integrating along vertical 
lines in the transition rate map, see Fig.~\ref{MAP1}. The function $\epsilon_g(t)$ 
has a complicated form with variable slope, which depends not only on the photon 
energy sweep $b$ but also on the pulse duration. The fact that $b$ cannot be 
immediately inferred from $d\epsilon_g(t)/dt$ is a combined effect of two factors: 
the FWHM of the EDC changes with time, and the broader the EDC the stronger it is 
affected by the energy dependence of the cross-section $S(\epsilon)$. In the present 
example the cross-section is higher at lower energies, see right panel of 
Fig.~\ref{f3}, so the broader the frequency distribution the larger the downward 
shift of the center of gravity. This produces a different effect on the spectra 
depending on the sign of the chirp, so the curves for opposite chirps in Fig.~\ref{f3} 
are not symmetric. At $t=D_{\textsc x}/2$ the curves of opposite chip converge, which 
follows from Eq.~(\ref{simple}) because the spectrum $I(\omega)$ of a linearly 
chirped pulse does not depend on the sign of chirp. 

The function $J(\epsilon,t)$ is the most fundamental quantity that contains 
information about the details of the ionization process. In the simple case
considered above the properties of the system enter via the energy dependent
cross-section $S(\epsilon)$, and its structure is seen to be reflected in
$J(\epsilon,t)$. At the same time, this means that in general the XUV pulse 
cannot be satisfactorily characterized by gross features of the resulting 
photoelectron spectrum evolution. In addition, the function $J(\epsilon,t)$ 
cannot be directly measured in an experiment, which makes the problem of 
disentangling the pulse characteristics from the properties of the system 
highly nontrivial. The only way to probe $J(\epsilon,t)$ is the laser streaking 
technique, which we consider in the next section.

\section{effect of laser field on lineshape}\label{LaserDressedNoChirp}
Let us consider photoelectrons emitted parallel to the electric field of a 
linearly polarized light. They are accelerated or decelerated depending 
on the sign of the momentum transfer $\Delta p(t) = (e/c)A(t)$ from the
laser field to an electron created at the time $t$. As a result, the spectrum 
shifts as a whole on the kinetic energy scale and distorts. One reason for 
the distortion is a finite width of the photoelectron spectrum: electrons 
with different momenta $p$ acquire different energies, so the spectrum 
broadens when accelerated and narrows when decelerated. Another source of 
distortion is the finite duration $D_{\textsc x}$ of the XUV pulse: electrons 
created at different moments $t$ are differently accelerated according to 
the current value of the momentum transfer $\Delta p(t)$.

\subsection{Initial time-momentum distribution}\label{LaserDressedNoChirp_A} 
In the absence of the laser field the momentum distribution of photoelectrons 
$J(p)$ after the XUV pulse is over can be expressed as the integral of the 
transition rate:
\begin{equation}\nonumber
J(p)=
\int\limits_{t_{\textsc x}-\frac{1}{2}D_{\textsc x}}
           ^{t_{\textsc x}+\frac{1}{2}D_{\textsc x}}r_{0}(p,t)\;dt.
\end{equation}
This formula can be generalized to the case of laser assisted photoemission by 
considering $J(p)$ as a projection of the initial time-momentum distribution 
onto the scale of final momenta $p$, which is the essence of the classical 
picture of the streaking measurement~\cite{Kie2004N}: 
\begin{equation}\label{classic}
J(p)=
\int\limits_{t_{\textsc x}-\frac{1}{2}D_{\textsc x}}
           ^{t_{\textsc x}+\frac{1}{2}D_{\textsc x}}r_{\rm ini}[p-\Delta p(t),t]\;dt.
\end{equation}
This equation means that the electron created at a moment $t$ in the state 
$p-\Delta p(t)$ is brought by the laser field to the final state $p$, and 
all such processes at different moments $t$ during the action of the XUV 
pulse are summed up incoherently.
\begin{figure}[b]    
\begin{minipage}{0.45\textwidth}
\includegraphics[width=\textwidth]{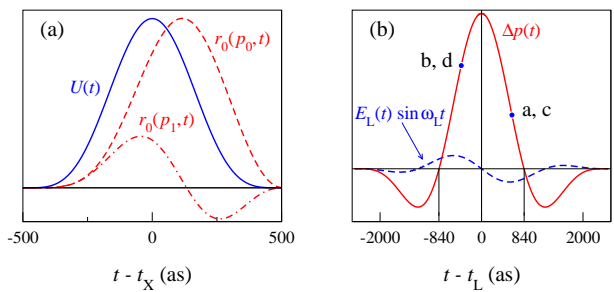}
\end{minipage}
\caption{\label{F4} (color online)
(a) Temporal intensity profile of the XUV pulse envelope $U(t)$, see 
    Eq.~(\ref{u_envelope}), (solid line) compared to the laser-free 
    transition rate functions $r_0(p,t)$ for the central momentum, 
    $p_0^2=\epsilon_{\rm i}+\omega_{\textsc x}-V_0$ (dashed line),
    and for a momentum far from the center $p_1$ (dot-dashed line).
(b) Electric field $E_{\textsc l}(t)$ of the laser pulse (dashed line) and the 
    transferred momentum $\Delta p(t)$. The circles indicate the release points
    $t_{\textsc x}$ of the XUV pulses studied in Fig.~\ref{F5}.     
}
\end{figure}                          

\begin{figure}[t]    
\begin{minipage}{0.45\textwidth}
\includegraphics[width=\textwidth]{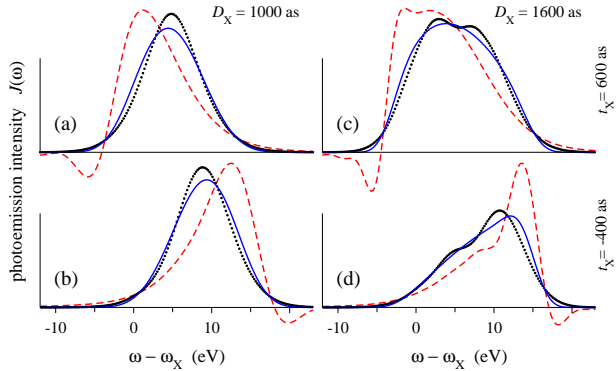}
\end{minipage}
\caption{\label{F5} (color online)
Streaked photoelectron EDCs for XUV pulse durations $D_{\textsc x}=1000$~as 
(a, b) and 1600~as (c, d) and release points $t_{\textsc x}=600$~as (a, c) 
and -400~as (b, d) indicated in Fig.~\ref{F4}(b). Dotted lines are the full
solution of the TDSE, and solid lines are obtained from the SSA equation 
(\ref{classic}) with the function $r_{\rm ini}$ from Eq.~(\ref{phemod}). Dashed 
curves result from using the laser-free function $r_{\rm 0}$ in Eq.~(\ref{classic}).
}
\end{figure}

Let us concentrate on the physical meaning of the function $r_{\rm ini}(p,t)$.
This function characterizes the XUV pulse, and at the same time it depends on 
the properties of the electronic system. To define this function requires a 
separation of the XUV and the laser pulses, which is, in general, not possible. 
For example, the time-momentum distribution in the absence of the laser field 
$r_0(p,t)$ cannot serve this purpose. First, even in the simplest case of a 
bandwidth limited pulse, for momenta far enough from the central momentum the 
function $r_0(p,t)$ becomes negative because the momentum distribution of 
photoelectrons narrows with time. This is illustrated by the dot-dashed line 
in Fig.~\ref{F4}(a), which shows two examples of $r_0(p,t)$ calculated by the 
TDSE. The implication for the laser assisted photoemission is shown in 
Fig.~\ref{F5}, which compares exact streaked spectra from the TDSE with those 
from the SSA. Although the contribution from negative values of $r_0$ is small 
it is not negligible, which is seen by the negative values of the spectral 
intensity in the dashed curves in Fig.~\ref{F5}. More important is that also 
around the central momentum, for which $r_0$ is everywhere positive [dashed 
line in Fig.~\ref{F4}(a)], it provides a rather inaccurate final distribution 
$J(p)$ (compare dotted and dashed lines in Fig.~\ref{F5}).

Instead of $r_0$ we can use in Eq.~(\ref{classic}) the separable ansatz 
\begin{equation}\label{phemod}
r_{\rm ini}(p,t) \rightarrow U(t) I(p) S(p)
\end{equation}
introduced in Ref.~\cite{KB07}, which turns out to perform much better, see 
solid lines in Fig.~\ref{F5}. In the phenomenological formula (\ref{phemod}) 
the explicit time dependence of the initial time-momentum distribution comes 
solely through the temporal intensity envelope $U(t)$ of the XUV pulse [see 
Eq.~(\ref{u_envelope})], which is shown by the solid line in Fig.~\ref{F4}(a). 
(Oscillations with the frequency $\omega_{\textsc x}$ are, thus, completely 
ignored.) In the presence of the laser field the functions $I$ and $S$ in 
Eq.~(\ref{classic}) also depend on time through the argument $p-\Delta p(t)$. 
The formula~(\ref{phemod}) establishes a simple connection between the structure 
of the XUV pulse and the streaked photoelectron spectrum, and it can be useful 
for the characterization of the pulse provided it reasonably simulates the true 
spectrum. It satisfies the two limiting cases: first, for an extremely short 
pulse, $\Delta p(t)$ can be considered constant, and 
Eqs.~(\ref{classic})--(\ref{phemod}) yield a shift of the spectrum as a whole. 
Second, in the absence of the laser field the equation~(\ref{simple}) is recovered.

Figure~\ref{F4}(a) compares laser-free functions $r_0(\epsilon,t)$ with the 
function $U(t)$ used in the ansatz (\ref{phemod}). At the Fermi golden rule 
energy $\epsilon_{\rm i}+\omega_{\textsc x}$ the true population profile $r_0$ 
is retarded by some 100~as with respect to the XUV intensity profile $U(t)$, 
which is sufficient to cause an error of more than 3~eV in the location of the 
spectral maximum. At the same time, equation (\ref{classic}) with the heuristic 
ansatz (\ref{phemod}) reproduces the spectra with a surprisingly high quality 
(compare dotted and solid lines in Fig.~\ref{F5}). Thus, in view of the sensitivity 
of the streaked spectra to small details of the function $r_{\rm ini}(\epsilon,t)$, 
the prescription (\ref{phemod}) appears to be a reasonable choice. The model
performs well for durations well exceeding a linear interval of the $\Delta p(t)$ 
function. It describes the dependence of the spectral width on release point and 
pulse duration, as well as the asymmetric shape of the streaked EDCs 
[see Fig.~\ref{F5}(d)].

\subsection{Effect of a rapid change of momentum transfer}\label{e_stretch}
A detailed discussion of the laser field induced line-shape variations of 
photoemission spectra has been presented in Ref.~\cite{KB07}. We shall now 
consider the possibility of inferring the XUV pulse duration from the measured 
streaked EDC. At a point where $\Delta p(t)$ rapidly changes with time, 
electrons created at different moments $t$ are differently accelerated 
according to the current value of $\Delta p(t)$, so the longer the pulse is, 
the stronger the EDC broadens. This ``temporal'' broadening is accompanied 
by the energy stretch of the spectrum due to a uniform shift along the $p$ 
axis (electrons with higher kinetic energies acquire larger energy from the 
laser field than those with lower energies). If the integral momentum transfer 
can be neglected [around $t-t_{\textsc l}=\pm 840$~as, see Fig.~\ref{F4}(b)] 
the latter effect is minimal. It has been suggested in Ref.~\cite{Kie2004N}
that in this case the temporal emission profile $U(t)$ is uniquely mapped into 
a spectral distribution of photoelectrons $J(\epsilon)$, and temporal information
can be retrieved from a single streak record.
\begin{figure}[b]    
\begin{minipage}{0.45\textwidth}
\includegraphics[width=\textwidth]{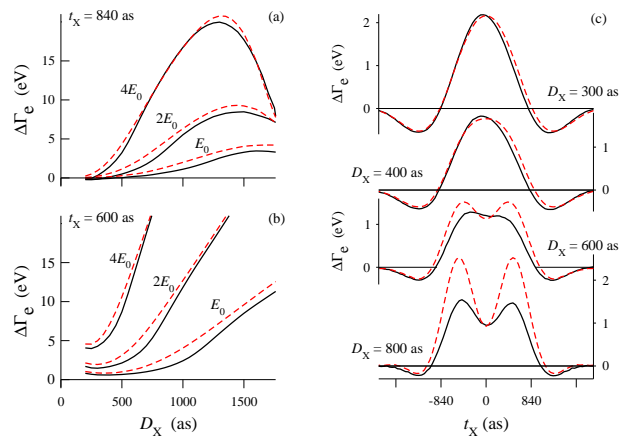}
\end{minipage}
\caption{\label{f6} (color online)
Graphs (a) and (b) show the dependence of the streaking broadening 
$\Delta\Gamma_{\rm e}$ on the duration $D_{\textsc x}$ of the XUV pump 
pulse for two release points $t_{\textsc x}=840$~as (a) and 600~as (b) 
by the full solution of the TDSE (solid lines) and by the SSA (dashed 
lines). The curves are presented for three values of the laser field 
amplitude: $E_{\textsc 0}=6\times 10^7$~V/cm, as well as $2E_{\textsc 0}$ 
and $4E_{\textsc 0}$. Graph (c) shows the dependence of $\Delta\Gamma_{\rm e}$ 
on release time for $D_{\textsc x}=300$, 400, 600, and 800~as.
}
\end{figure}                       

Owing to the finite spectral width of the XUV pulse the streaking broadening 
$\Delta\Gamma_{\rm e}$ (i.e., the difference between the FWHM of the 
photoelectron spectra in the presence and in the absence of the laser field)
depends on the pulse duration $D_{\textsc x}$ in a complicated manner. 
Figures~\ref{f6}(a,b) show the dependence of $\Delta\Gamma_{\rm e}$ on $D_{\textsc x}$ 
for two release points $t_{\textsc x}$ and three laser field intensities. 
Up to durations well exceeding the linear interval of $\Delta p(t)$ the 
streaking broadening steadily increases with $D_{\textsc x}$ while the 
FWHM of the field-free spectrum steadily decreases. As a result, two 
different temporal profiles $U(t)$ may give almost identical streaked spectra.

The duration can, however, be immediately inferred from $\Delta\Gamma_{\rm e}$,
and the SSA gives a reasonable approximation over a wide range of laser intensities. 
For small $D_{\textsc x}$ the width of the non-streaked spectrum increases, and 
the absolute values of $\Delta p(t)$ start playing a more important role: the 
energy stretch (which varies in time) interferes with the temporal broadening. 
At the release point $t_{\textsc x}=600$~as the integral momentum transfer is much 
larger, so the energy stretch at small $D_{\textsc x}$ leads to the presence of a 
minimum in the $\Delta\Gamma_{\rm e}(D_{\textsc x})$ curve.

At short durations, the streaking broadening not only becomes small 
at the point of zero momentum transfer, but also the sensitivity of 
$\Delta\Gamma_{\rm e}$ to the release point increases, as illustrated 
by Fig.~\ref{f6}(c). This means that this point is less favorable for 
the measurement of $D_{\textsc x}$ than the measurement at the point 
of maximal momentum transfer. Figure~\ref{f6}(c) suggests that 
$\Delta\Gamma_{\rm e}$ measured at $t_{\textsc x}=0$ (maximum $\Delta p$) 
and $t_{\textsc x}=\pm 1250$~as (minimum), as well as the $t_{\textsc x}$
location of the extrema would characterize the XUV pulse more accurately.

Figure~\ref{f6} shows that the results by the SSA~(\ref{classic})--(\ref{phemod}) 
closely follow the exact solutions (except for the overestimated maximal broadening 
for longer pulses), and  their quality does not deteriorate with increasing the 
streaking field strength.

\section{Extension of SSA to chirped pulses}\label{LaserDressedChirpedPulse}
We now generalize the semiclassical approach to chirped pulses. We again rely on 
Eq.~(\ref{classic}) to separate the pump and probe pulses, but Eq.~(\ref{phemod}) 
is not applicable now because it cannot discriminate between positive and negative 
chirp. Indeed, the pump pulse enters through its spectrum $I(\omega)$ and time 
envelope $U(t)$, while both functions do not depend on the sign of the chirp
rate $b$ [defined in Eq.~(\ref{chirp_formula})]. In order to include the photon 
energy sweep in our model we must ascribe physical meaning to the formal notion 
of instantaneous frequency $\omega_{\textsc x}(t)$. Thereby the spectral intensity 
distribution $I(\omega)$ in Eq.~(\ref{phemod}) will acquire an {\em explicit} 
time dependence. To achieve this we introduce an `instantaneous' spectral intensity 
distribution $g(\omega,t)$. In order to satisfy the basic equation (\ref{simple}) 
in the absence of the laser field, this function should obey the relation 
\begin{equation}\label{relation}
I(\omega) = \int\limits_{-\infty}^{+\infty} g(\omega,t) U(t)\, dt.
\end{equation}
For a linear chirp and a temporal intensity distribution of Gaussian 
form, $U(t)\sim\exp[-(16\ln 2)\,t^2/D_{\textsc x}^2]$, the spectral 
intensity distribution $I(\omega)$ is again a Gaussian with FWHM 
$\Gamma_{\textsc x}=[64(\ln 2)^2+D_{\textsc x}^4b^2]^{\frac{1}{2}}/D_{\textsc x}$,
and equation (\ref{relation}) has the explicit solution: 
\begin{equation}\label{moving_Gauss}
g(\omega,t)=\exp-\frac{(\omega-\omega_{\textsc x}^0-bt)^2}{\gamma^2/4\ln 2}, \quad
\gamma^2 =\Gamma_{\textsc x}^2-\frac{D_{\textsc x}^2b^2}{4}.
\end{equation}
Although in our case the function $U(t)$ deviates from Gaussian we shall
employ this form of $g(\omega,t)$ and the requirement that the convolution
(\ref{relation}) yield the FWHM $\Gamma_{\textsc x}$ of the actual spectral
distribution $I(\omega)$. 

The semiclassical superposition (\ref{classic}) with the ansatz
\begin{equation}\label{newmod}
r_{\rm ini}(p,t) = U(t) g(p,t) S(p)
\end{equation}
for the initial momentum distribution takes into account both the temporal
and the spectral structure of the pump pulse. Thus, from the point of view
of photoemission, the pump pulse is thought to produce at the time point $t$
transitions with the spectral intensity distribution proportional to $g(\omega,t)$
[centered at $\omega_{\textsc x}(t)$] and the maximal intensity proportional to
$U(t)$. In the limit of zero chirp, according to the requirement (\ref{relation}),
the function $g(\omega,t)$ becomes time independent and reduces to $I(\omega)$,
which leads back to Eq.~(\ref{phemod}). Note that in contrast to the pulselet 
representation of Ref.~\cite{KK06}, the function $g(\omega,t)$ does not contain
phase, and its width is much smaller than the spectral width of a pulselet. 

\begin{figure}[b]    
\begin{minipage}{0.4\textwidth}
\includegraphics[width=\textwidth]{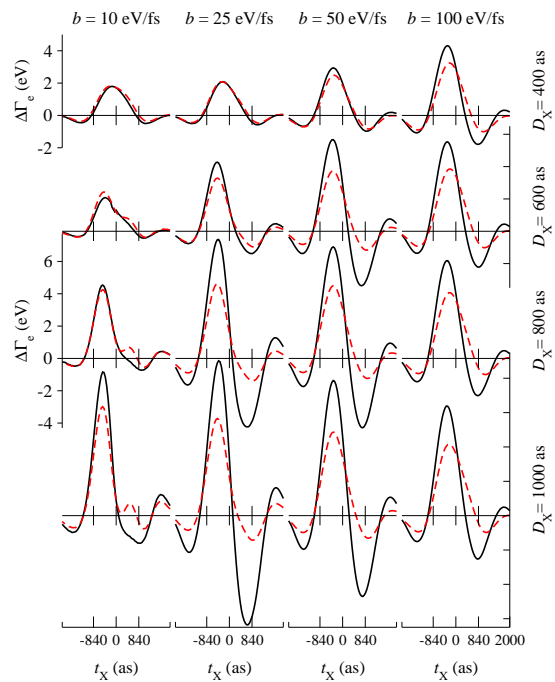}
\end{minipage}
\caption{\label{f7} (color online)
Dependence of the streaking broadening $\Delta\Gamma_{\rm e}$ of the 
photoelectron spectra on the release point $t_{\textsc x}$ for four 
values of chirp rate $b=10$, 25, 50, and 100~eV/fs and four pulse 
durations $D_{\textsc x}=400$, 600, 800, and 1000~as. Solid lines 
are the exact solution of the TDSE and (red) dashed lines are obtained 
from the SSA equations (\ref{classic}) and~(\ref{newmod}). 
}
\end{figure}                       
Let us see how well the SSA (\ref{classic}) with the initial distribution 
(\ref{newmod}) describes the dependence of the streaking broadening 
$\Delta\Gamma_{\rm e}$ on the chirp rate $b$ of the pump pulse. For $b=0$ the 
broadening $\Delta\Gamma_{\rm e}$ as a function of $t_{\textsc x}$ is practically 
symmetric with respect to $t_{\textsc x}=0$, see Fig.~\ref{f6}(c). 
For long pulses the ``temporal'' broadening mechanism dominates, and the curve  
$\Delta\Gamma_{\rm e}(t_{\textsc x})$ has two distinct maxima (which merge into a 
broad one at $t_{\textsc x}=0$ when the duration decreases). The symmetric shape 
is completely distorted already at small chirp rates, see Fig.~\ref{f7}. One of 
the maxima is strongly damped and the other one is enhanced, so the shape of the 
curve provides a signature of the temporal structure of the XUV pulse.

At a zero momentum transfer point the photoelectron spectrum excited by a 
positively chirped pulse is stretched by the positive and compressed by the 
negative laser field. This means that from the measured $\Delta\Gamma_{\rm e}$ 
at the two $t_{\textsc x}$ points one can infer both the duration and the chirp 
of the pump pulse~\cite{Kie2004N}. Indeed, Fig.~\ref{f7} shows that both the 
broadening and the compression steadily increase with the pulse duration. 
However, the dependence on chirp rate is more complicated: for a fixed 
$D_{\textsc x}$ the amplitude of $\Delta\Gamma_{\rm e}$ variations first 
increases with the chirp rate, and at larger $b$ it starts decreasing. This 
non-monotonic behavior is caused by the interference of the energy stretch 
mechanism discussed in Sec.~\ref{e_stretch}: over the whole XUV pulse duration 
electrons with a wide spread of energies are created, which are differently 
affected by the laser field depending on their energies. At higher chirp
rates the spectral width $\Gamma_{\textsc x}$ of the pump pulse increases, which 
blurs the signature of its temporal structure. In addition, with increasing 
$\Gamma_{\textsc x}$ the energy dependence of the cross-section [factor $S(p)$ 
in Eq.~(\ref{newmod})] becomes more and more important, whereby the laser-free 
photoelectron spectral width $\Gamma_{\rm e}^0$ increasingly strongly deviates 
from $\Gamma_{\textsc x}$ (see also a discussion in Ref.~\cite{KB07} and the 
comment on Fig.~\ref{f3} in the end of Sec.~\ref{LaserFreeCase}). 

The SSA is found to well reproduce the spectral width variations for 
$\eta=bD_{\textsc x}/\Gamma_{\textsc x}^0<1.5$, which covers a wide range of
experimentally interesting pulses. For longer pulses or larger chirp rates it 
gives only a qualitative picture, see Fig.~\ref{f7}. The model systematically 
underestimates the amplitude of the $\Delta\Gamma_{\rm e}$ oscillations, with the 
largest discrepancy occurring when the energy sweep causes a compression of the 
spectrum. This has to be expected as the quantum interference between different 
processes that bring the excited electron to a given final state becomes most 
important in this case. The broadening of the spectra is reproduced much better; 
in particular, the time point at which the spectral width is maximal is predicted 
rather accurately. The numerical experiments, thus, suggest that the limitations 
of the SSA lie in treating the interaction with the laser field, i.e., in 
Eq.~(\ref{classic}), whereas the heuristic description of the energy sweep by the 
instantaneous intensity distribution in Eq.~(\ref{newmod}) is quite reasonable.

In view of the complicated interplay between different broadening mechanisms 
and because the pulse width $\Gamma_{\textsc x}$ is a non-linear function of 
both $D_{\textsc x}$ and $b$ the question arises whether a chirped pulse can be 
characterized by extracting only the gross parameter $\Delta\Gamma_{\rm e}$ 
from the streaking measurements. A more general but complicated procedure is 
to measure a complete spectrogram $J(\epsilon,t_{\textsc x})$ and decipher it 
with a FROG algorithm (frequency-resolved optical gating), which under certain 
conditions allows to reconstruct both the intensity profile and the phase of 
the pulse~\cite{GGY08}. A physically more transparent way is to establish a 
relation between gross features of the input pulse and the output spectrum 
and consider the release points at which the output is most sensitive to input. 
Following Ref.~\cite{Kie2004N}, we shall consider zero momentum transfer points.

\begin{figure}[t]    
\begin{minipage}{0.48\textwidth}
\includegraphics[width=\textwidth]{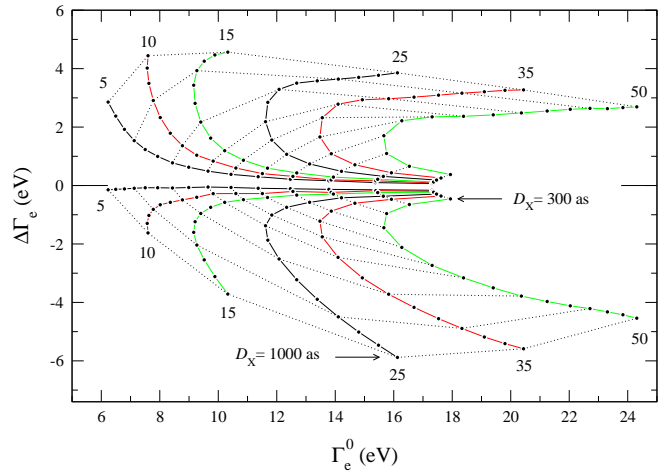}
\end{minipage}
\caption{\label{f8} (color online)
Dependence of the laser-free spectral FWHM $\Gamma_{\rm e}^0$ and streaking
broadening $\Delta\Gamma_{\rm e}$ on the duration $D_{\textsc x}$ and chirp 
$b$. Positive $\Delta\Gamma_{\rm e}$ correspond to the release point 
$t_{\textsc x}=-840$~as and negative to $t_{\textsc x}=840$~as. The 
$b=\rm const$ lines (solid lines) are labelled by chirp rate in the
units of eV/fs. The dots in the lines are placed at equal intervals 
of 50~as between $D_{\textsc x}=300$ and 1000~as. Points of equal duration
are connected by dotted lines.
}
\end{figure}

In order to determine the temporal parameters $b$ and $D_{\textsc x}$ from the 
spectral characteristics $\Gamma_{\rm e}^0$ and $\Delta\Gamma_{\rm e}$ the input 
values must be presented in the output coordinates. The map in Fig.~\ref{f8} 
shows $b=\rm const$ lines parametrized by $D_{\textsc x}$ for the two zero momentum 
transfer points, see Fig.~\ref{F4}(b). The data are obtained with the exact TDSE.
According to our calculations, neither the $b=\rm const$ lines nor the 
$D_{\textsc x}=\rm const$ ones intersect, which means that, in principle, both $b$ 
and $D_{\textsc x}$ can be determined from a laser-free EDC and a single streaked 
measurement (for one of the $\Delta p=0$ points). The sensitivity of output to 
input strongly varies over the map: the uncertainty of $D_{\textsc x}$ increases 
with $D_{\textsc x}$, especially for small chirp rates, and at short durations the 
uncertainty of $b$ increases. The complicated structure of the map implies that 
solution of the inverse problem requires a careful consideration of the photoexcitation 
process, as well as very accurate measurement of the EDCs.

\section{summary}
The main result of this work is the extension of the well known factorization
$J(\epsilon)\sim I(\epsilon)S(\epsilon)$ of the photoelectron spectrum by 
a single short pulse to laser assisted photoemission [as expressed by 
Eqs.~(\ref{classic}), (\ref{phemod}), and (\ref{newmod})]. This simple relation 
tells us that the laser free photoelectron spectrum is determined by the spectrum 
of the exciting pulse but does not depend upon its temporal structure. In the 
streaking measurements it is just the temporal structure which is probed, and 
we have shown that the resulting spectrum depends on the intensity envelope, 
while the phase structure of the pump pulse is blurred. In the case of a Fourier 
limited pulse it enters only indirectly, as the relation between the temporal 
and the spectral extent of the pulse. In the case of a chirped pulse it is 
insufficient to know the time envelope and the spectrum. We have shown that
this more complicated temporal structure can to a good approximation be treated 
in spirit of the formal instantaneous frequency by introducing an instantaneous 
spectral distribution and again neglecting the actual phase relations. 

We have applied these considerations to calculating streaked spectra within 
the semiclassical superposition approximation. Although the concept of initial 
time-momentum distribution $r_{\rm ini}(p,t)$ lacks a rigorous physical meaning 
a function with the required properties can be constructed, which is capable of 
describing the lineshapes of streaked spectra for chirped pump pulses. In addition 
to the temporal structure of the pulse it takes into account its spectral extent, 
i.e., the fact that during the action of the pulse electrons are created in all 
states but with different rate. Both the temporal and the spectral factors are 
important for the shape of the streaked EDC, and the photon energy sweep, which 
both complicates the temporal structure and broadens the spectrum, provides a 
severe test of the SSA model. The model yields quantitatively correct results 
for a wide range of parameters and provides a useful basis for the understanding 
of the role of pulse parameters and properties of the electronic system. The 
latter are included through the energy dependent photoemission cross-section, 
which was found to strongly affect the lineshapes. The property of the model 
to decouple the semiclassical calculation of the time evolution of the system 
from the quantum-mechanical calculation of the dipole matrix elements makes it 
attractive for use for more complicated many-body systems, for which a consistent 
quantum-mechanical calculation is currently computationally infeasible: one can 
reasonably suppose that the present formalism would remain valid with a more 
advanced description of the excitation, see, e.g. Ref.~\cite{HBBB} and references 
therein.

The main limitations of the model naturally come from its neglect of the 
quantum interference that accompanies the acceleration of the wave packet 
by the laser field. In the case considered this led to underestimated 
broadening and compression for sufficiently large pulse durations and 
chirp rates, but did not affect the qualitative picture.

The phenomenological function $r_{\rm ini}$ suggested in this paper differs from 
the time-momentum distribution $r_{0}$ in the absence of the laser field in the 
two essential features: it discards the retardation of the final state population 
with respect to the pulse intensity profile and avoids negative values of the 
population rate. Thereby the function $r_{\rm ini}$ provides the most direct 
connection between the input pulse and the output EDC, and, moreover, the two 
aspects are found instrumental in reproducing the shape of the streaked EDCs 
within the SSA. This means, however, that in solving the inverse problem, i.e., 
obtaining information about the pump process from streaking measurements some 
essential features of the process may be lost. 

We have shown that in the simplest case of a linear chirp the relation between the 
input pulse and output spectrum can be presented on a map that establishes a one to 
one correspondence between gross features of the input and the output. This means 
that provided a reliable theory of photoemission is available and the experimental 
resolution is high the pump pulse can be characterized by a few streak records.

\vfill

\begin{acknowledgments}
The authors gratefully acknowledge fruitful discussions with N.M.~Kabachnik, 
A.K.~Kazansky, E.V.~Chulkov and P.M.~Echenique. The work was supported by Ikerbasque 
(Basque Foundation for Science) and by the Innovationsfond Schleswig-Holstein.
\end{acknowledgments}

\appendix*\section{}
We shall show that the simple relation $J(\epsilon)\sim I(\epsilon)S(\epsilon)$
is obtained in the perturbational limit $eA(t)/c \ll p$ of the time-dependent
Schr\"odinger equation if one neglects the non-diagonal dipole matrix elements 
between final states. In order to avoid the necessity of treating the final 
states as plane waves we consider a bound system (e.g., a quasi-continuum as in 
Fig.~\ref{box}) with the unperturbed Hamiltonian $\hat H_0=-\Delta+V(x)$ and a 
time-dependent perturbation $E(t)\,\hat x$. Let us write the Schr\"odinger equation 
\begin{equation}\label{TDSE1}
i\dot\psi(t)=\left[\,\hat H_0 + E(t)\,\hat x\,\right]\psi(t)
\end{equation}
in terms of the discrete spectrum of the eigenstates $|\,j\,\rangle$ of 
the unperturbed operator, $\hat H_0|\,j\,\rangle=\epsilon_j|\,j\,\rangle$. 
The solution consists in finding the expansion coefficients $a_j(t)$:
\begin{equation}
|\,\psi(t)\,\rangle=\sum\limits_{j=0}^{\infty}a_j(t)\,|\,j\,\rangle.
\end{equation}
The time-dependent Schr\"odinger equation then reads
\begin{equation}\label{TDSE}
i\sum\dot a_j(t)\,|\,j\,\rangle=\left[\,\hat H_0 + E(t)\,\hat x\,\right]
 \sum     a_j(t)\,|\,j\,\rangle\,.
\end{equation}

Let $j=0$ be the initial state, i.e., $a_0(0)=1$ and  $a_j(0)=0$ for $j\ne 0$. 
It can be shown that for a sufficiently weak potential $V(x)$ the non-diagonal
matrix elements of the position operator $\hat x$ are much smaller than the
diagonal ones. Let us retain in Eq.~(\ref{TDSE}) only the matrix elements 
$d_{jj}=\langle\, j\,|\,\hat x\,|\,j\,\rangle$ and 
$d_{j0}=\langle\, j\,|\,\hat x\,|\,0\,\rangle$. Note that this approximation 
does not depend on the strength of the perturbing field $E(t)$. Its validity 
depends only on the properties of the system. Then the equations for different 
$j$ separate:
\begin{equation}\label{separate}
i\dot a_j(t)=a_0(t)E(t)d_{j0} + a_j(t)[\epsilon_j+d_{jj}E(t)].
\end{equation}
Let us write $a_j(t)$ as a product $a_j(t) = \exp[-i\epsilon_j t] \xi_j(t)$.
Then the equation for $\xi_j(t)$ is
\begin{equation}\label{forxi}
i\dot\xi_j(t)= \exp[i\omega_{j0}t]\xi_0(t)E(t)d_{j0} + \xi_j(t)E(t)d_{jj}.
\end{equation}
Here $\omega_{j0}=\epsilon_j-\epsilon_0$. The solution of Eq.~(\ref{forxi}) is
\begin{equation}
\begin{split}
\xi_j(t)=
-i\int\limits_{-\infty}^{t}dt'\,
\xi_0(t')E(t')\,d_{j0}\,\exp{[i\omega_{j0}t']}\\
\times\exp{\left[id_{jj}\int\limits_t^{t'}E(t'')\,dt''\right]},
\end{split}
\end{equation}
or, in terms of vector potential $A(t)$,
\begin{equation}\label{solution}
\begin{split}\xi_j(t)=-i\int\limits_{-\infty}^{t}dt'\,\xi_0(t')E(t')
\,d_{j0}\,\exp[i\omega_{j0}t']\,\\
\times\exp[id_{jj}(A(t')-A(t))].
\end{split}
\end{equation}
At this point we introduce the requirement that the perturbation be small. Its 
implication is twofold: first, for $A(t)\to 0$ the initial state can be considered 
unaffected by the electric field, so that $\xi_0(t')=1$. Second, the phase 
$d_{jj}(A(t')-A(t))$ can be neglected compared to $\omega_{j0}t'$. Then $\xi_j(t)$ 
reduces to the Fourier transform $E(\omega)$ of the XUV pulse multiplied by the 
dipole matrix element $d_{j0}$:
\begin{equation}\label{ultimate}
\xi_j(t)=-i\,d_{j0}\,\int\limits_{-\infty}^{t}dt'\,E(t')\,\exp[i\omega_{j0}t'].
\end{equation}

The photoelectron spectrum after the pulse is over 
$J(\omega_{j0})=|\xi_j(D_{\textsc x}/2)|^2$ is then a product of the 
photoemission cross-section $S(\omega_{j0})=|d_{j0}|^2$ and the 
intensity spectrum $I(\omega_{j0})=|E(\omega_{j0})|^2$. (Of course, 
Eq.~(\ref{ultimate}) can be applied to any $t$, i.e., it describes 
the evolution of the spectrum, too.) This result is also obtained in 
the perturbational limit of the equations of Ref.~\cite{LBI94}, but 
in the present derivation the assumption of plane-wave-like final states 
is recast as the approximation of vanishing non-diagonal elements of the 
perturbation between final states.
\end{document}